\documentclass[useAMS,usenatbib,usegraphicx]{mn2e}
\usepackage{times, deluxetable}
\usepackage{color}

\newcommand{\apj}{ApJ}           
           
\newcommand{\apjs}{ApJS}           
\newcommand{\mnras}{MNRAS}       
\newcommand{\nat}{Nature}
\newcommand{\aap}{A\&A}

\newcommand{\pasp}{PASP}
\newcommand{\sauron}{\texttt{SAURON}}
\newcommand{\atl}{ATLAS$^{\rm 3D}$}
\newcommand{\ppxf}{p\textsc{pxf}}
\newcommand{\kms}{\hbox{$\rm km\, \rm s^{-1}$}}
\newcommand{\re}{\hbox{$R_{\rm e}$}}

\newcommand{\epse}{\hbox{$\epsilon_{\rm e}$}}
\newcommand{\lambdae}{\hbox{$\lambda_R(R_{\rm e})$}}
\newcommand{\fsr}{\hbox{$f_{\rm SR}$}}
\newcommand{\MK}{\hbox{$\rm M_{\rm K}$}}

\newcommand{\refsec}[1]{Section~\ref{#1}}
\newcommand{\reffig}[1]{Fig.~\ref{#1}}

\newcommand{\reftab}[1]{Table \ref{#1}}
\newcommand{\text}[1]{\textrm{#1}}
\newcommand{\Angst}{$\mbox\AA$}

\title[A FLAMES/GIRAFFE
IFS study of galaxies in Abell 1689 at z=0.183]{Fast and slow rotators in the densest environments: a FLAMES/GIRAFFE
IFS study of galaxies in Abell 1689 at z=0.183}

\author[F.~D'Eugenio et al.]
{F. D'Eugenio$^1$\thanks{E-mail: Francesco.D'Eugenio@physics.ox.ac.uk}, 
R. C. W. Houghton$^1$, 
Roger L. Davies$^1$ 
and E. Dalla Bont\`a$^{2,3}$\\
$^1$Sub-department of Astrophysics, Department of Physics, University of
Oxford, Denys Wilkinson Building, Keble Road, Oxford OX1 3RH\\
$^2$Dipartimento di Fisica e Astronomia ``G. Galilei'', Universit\`a degli
Studi di Padova, Vicolo dell'Osservatorio 3, I-35122, Padova, Italy\\
$^3$INAF Osservatorio Astronomico di Padova, Vicolo dell'Osservatorio 5,
I-35122, Padova, Italy
}

\pagerange{\pageref{firstpage}--\pageref{lastpage}}
\pubyear{2011}

\begin{document}

\maketitle

\label{firstpage}

\begin{abstract}
  We present FLAMES/GIRAFFE integral field spectroscopy of 30 galaxies in the massive cluster Abell 1689 at z = 0.183. Conducting
an analysis similar to that of \atl, we extend the baseline of the
kinematic morphology-density relation by an order of magnitude in
projected density and
show that it is possible to use existing instruments to identify slow
and fast rotators beyond the local Universe. We find $4.5 \pm 1.0$ slow rotators
with a distribution in magnitude similar to those in the Virgo cluster.
The overall slow rotator fraction of our Abell 1689 sample is
$0.15 \pm 0.03$,
the same as in Virgo using our selection criteria. This suggests
that the fraction of slow rotators in a cluster is not strongly dependent on
its density. However, within Abell 1689, we find that the fraction of
slow rotators increases towards the centre, as was also found in
the Virgo cluster.
\end{abstract}


\section{Introduction}

\subsection{Galaxies and environment density}

  Early type galaxies (ETGs), despite having masses and luminosities that span
several orders of magnitude, obey a number of tight phenomenological laws.
These, collectively known as ``scaling relations'', include the color-magnitude
diagram \citep[CMD:][]{baum1959, visvanathan1977, sandage1978a, sandage1978b},
the color-$\sigma$ and $Mg-\sigma$ \citep{burstein1984, bender1993} relations and the fundamental plane
\citep{dressler1987, djorgovski1987}. With
their remarkably small scatter, they impose strong constraints on the structure
and evolution of ETGs.
Morphologically, ETGs are either classified as ellipticals (Es) or lenticular
(S0) galaxies. In the late 1980s, based
on new and more accurate spectroscopy, Es were divided into two groups:
pressure supported and rotation supported \citep{bender1989}.
It is thus particularly interesting to investigate how such dynamically
distinct systems formed and evolved while still obeying the very same scaling
relations.
Environment certainly plays a major role in galaxy evolution, as witnessed by the morphology-density relation
\citep{dressler1980}: systems in denser surroundings are more likely to be ETGs.

The advent of integral field spectroscopy (IFS) has brought a wealth of
information to the field. The \sauron\ survey
discovered the existence of two kinematically distinct classes of ETGs, slow
and fast rotators \citep[SR and FR,][]{emsellem2007, cappellari2007}.
The former are systems with little to no rotation, often exhibiting
kinematically decoupled cores and misalignment between kinematics and
photometry. The latter are flattened systems, compatible with rotational
symmetry, where ordered, large scale rotation is important for the gravitational
equilibrium.
While overlapping with the existing dichotomy among ETGs, the new classification
crucially crosses the boundary between Es and S0s, in that FRs populate both
morphological classes. Indeed \atl\ \citep[the volume limited follow-up survey to
\sauron, ][]{cappellari2011a, emsellem2011}, found that many morphological Es are
FRs. They suggest a new classification paradigm based
on kinematics rather than morphology \citep{cappellari2011b}.\\

\atl\ also presented the kinematic morphology-density relation (kT-$\Sigma$),
linking the fraction of SRs (\fsr) with the local number density
of galaxies. \fsr\ is insensitive to environment density over 5 orders of
magnitude, with a sharp increase observed only in the inner core of the Virgo
cluster. \citet{cappellari2011b} conclude by
asking what would be measured in the denser environments beyond the local
Universe: does the fraction of slow rotators increase further or does it stay
constant?

Addressing this question would give further insight on the processes that drive
galaxy formation and evolution, and is indeed the goal of this work.

\subsection{This study}

  We exploited the unique capabilites of the FLAMES/GIRAFFE multiplexed
integral field spectrograph \citep{pasquini2002} at the
Very Large Telescope (VLT) to investigate internal kinematics of galaxies in the densest environment, so to extend the density baseline of the kT-$\Sigma$ relation. After
describing the observations in \refsec{sec:observations}, we present the data
reduction and analysis in \refsec{sec:data}. The results are
presented in \refsec{sec:results}, followed by their discussion in
\refsec{sec:discussion} and a summary in \refsec{sec:summary}.

\section[]{Observations}\label{sec:observations}

\subsection[]{Sample selection}\label{sec:sample.selection}

Abell 1689 is a massive galaxy cluster at redshift $z = 0.183$
\citep{struble1999}. Its
regular, concentric X-ray contours suggest it is a relaxed system
\citep{lemze2008}. An X-ray
luminosity of $\rm L_{\rm X} = 20.74 \times 10^{44}\, erg\, s^{-1}$ makes it
considerably more luminous than Coma, which has $\rm L_{\rm X} = 7.21 \times
10^{44}\, erg\, s^{-1}$ \citep{ebeling1996} and Virgo $\rm L_{\rm X} = 8.3
\times 10^{43}\, erg\, s^{-1}$ \citep{bohringer1994}. Assuming 7-yr
\textit{Wilkinson Microwave Anisotropy Probe} Cosmology
\citep[][$\Omega_m = 0.27, \Omega_\Lambda = 0.73, h_0 = 0.71$]{komatsu2011} its
comoving distance is 741 Mpc, giving 1$''$ per 3.0 kpc, so that GIRAFFE
deployable integral field units (see below) sample up to 1 \re\ for most galaxies.
GIRAFFE permits the observer to target 15 objects simultaneously and
we chose to target 30 galaxies as a compromise between sample size and
integration time. Our selection was based on a catalogue from
\citet{halkola2006}, and in order to gain the maximum possible
signal-to-noise ratio (SNR), we initially selected the 30 ETGs with the highest
surface brightness within \re\ (including the brightest cluster galaxy). This sample was then
subject to two practical constraints.
We needed all of our targets to have high resolution HST imaging, which limited
our choice to candidates in the innermost regions of the cluster.
Physical constraints from the instrument (see \refsec{sec:data:subsec:instrument})
ruled out some targets in the most crowded regions, forcing us to re-select
from a reserve list.
This left us with 29 galaxies inside the HST field of view and one outside
(galaxy 20).

\subsection[]{Archival data.}

We used F625W band imaging from the HST Advanced Camera for Surveys, combined
with \textit{g$'$} and \textit{r$'$} band GEMINI imaging. See
\citet{houghton2012} for a thorough description of these data.

\subsection{VLT data}\label{sec:data:subsec:instrument}

We present new data taken with the FLAMES/GIRAFFE spectrometer at the VLT Unit
Telescope 2. The
L612W filter gives a resolution R of 11800 (the
minimum allowed on the instrument) with a wavelength range of
5732-6515 \Angst\ (4858-5521 $\mbox{\AA}$ in the rest frame), which includes
prominent absorption features of old stellar populations (for comparison,
\sauron\ has a wavelength range of 4800-5380 \Angst). The observations were carried out between
24 May 2009 and 29 May 2009, as detailed in Table \ref{tab:obs.table}, which also
contains the observing conditions.
The instrument provides 30 independent integral
field units (IFUs), deployable anywhere on the focal plane. These are arranged
in two positioner plates, each hosting 15; each deployable IFU is positioned
by a magnetic button, with an accuracy better than $0.08''$ and contains an
array of 20 square microlenses, each with a side of $0.52''$ on the sky. They
are arranged in 4 rows of 6 (with 4 ``dead'' corners) for a total field of view
of $3'' \, \times 2''$. Each lenslet is then connected to the spectrometer with
a dedicated optical fibre. Alongside the 15 IFUs, each positioner plate also
houses 15 sky fibres. These are fully deployable just like the former
but carry only one lenslet.

\begin{table}
  \centering
  \caption[]{\label{tab:obs.table}A summary of the VLT/FLAMES GIRAFFE/IFU
    spectroscopy. The seeing was measured on site using telescope guide stars.}
  \begin{tabular}{lccccc}
    \hline
    \bf{Frame} & \bf{Plate} & \bf{Date} & night &\bf{Time} & \bf{Seeing} \\
    &  &  D/M/Y & & min & arcsec \\
    \hline
    \hline
   1 & 1 & 24 May 2009 & 1 & 120 & 0.60 \\
   2 & 1 & 25 May 2009 & 1 & 120 & 0.60 \\
   3 & 1 & 25 May 2009 & 2 & 120 & 0.60 \\
   4 & 1 & 26 May 2009 & 2 & 120 & 0.60 \\
   5 & 1 & 27 May 2009 & 3 & 120 & 0.60 \\
    \hline
   6 & 2 & 25 May 2009 & 1 & 120 & 0.50 \\
   7 & 2 & 26 May 2009 & 3 & 120 & 0.60 \\
   8 & 2 & 27 May 2009 & 3 & 120 & 0.60 \\
   9 & 2 & 27 May 2009 & 4 & 120 & 0.80 \\
  10 & 2 & 28 May 2009 & 4 & 120 & 0.65 \\
    \hline
  \end{tabular}
\end{table}

Since the magnetic buttons are larger ($10''$) than the IFU field of view,
they cannot be deployed closer than a minimum distance of $11''$ thus
constraining the sample selection: galaxies closer than $11''$ on the sky must
be allocated on different plates, if at all. As a result, some targets lying in the most
crowded regions of the cluster were omitted. We proceeded to divide the sample
in two equal sets, with galaxies numbers 1 to 15 assigned to plate 1 and
galaxies numbers 16 to 30 to plate 2. Each plate was exposed 5 times for 2
hours, for a total of 10 hours exposure time per galaxy.

We remark that, as detailed in \reftab{tab:obs.table}, the seeing was
comparable to the size of the lenslets ($0.52''$). This reduces the correlation
between adjacent spaxels.

\clearpage
\begin{table*}
\centering
\setlength{\tabcolsep}{4pt}
\caption[]{Our sample of 30 bright galaxies in Abell 1689.}\label{tab:sample}
\begin{tabular}{ccccccccccccccc}
\hline
\hline
 Galaxy  &  Halkola  &  RA  &  DEC  &
 $M_K$  &  $R_{\rm e}$  &  Q  &
 $\epsilon_{\rm e}$  &  PA  &
 p(SR)  &  $\lambda_R(IFU)$  &
 $\log\Sigma_3$  \\
   &    &  (deg)  &  (deg)  &
 (mag)  &  (arcsec)  &  (1-3)  &    &
   &  (0-1)  &  &
 (Mpc$^{-2}$)  \\
 (1)  &  (2)  &  (3)  &  (4) &
 (5)  &  (6)  &  (7)  &  (8) &
 (9)  &  (10)  &  (11)  &  (12) \\
\hline
1 & 17 & 197.94862 & -1.15575 & -23.28 & 0.65 $\pm$ 0.29 & 2 &	0.286 & 20 & 0.02 & 0.412 $\pm$ 0.092 & 2.381 \\
2 & 3 & 197.94831 & -1.14250 & -23.16 & 1.27 $\pm$ 0.57 & 1 &	0.583 & 163 & 0.00 & 0.585 $\pm$ 0.053 & 2.408 \\
3 & 12 & 197.95400 & -1.13998 & -23.57 & 0.71 $\pm$ 0.32 & 2 &	0.363 & 47 & 0.00 & 0.611 $\pm$ 0.087 & 2.279 \\
4 & 34 & 197.96433 & -1.14525 & -22.58 & 1.15 $\pm$ 0.51 & 2 &	0.032 & 235 & 0.00 & 0.455 $\pm$ 0.105 & 2.333 \\
5 & 22 & 197.96239 & -1.15608 & -24.97 & 2.84 $\pm$ 1.24 & 2 &	0.219 & 113 & 0.00 & 0.252 $\pm$ 0.022 & 2.928 \\
6 & 47 & 197.96764 & -1.16323 & -25.41 & 2.39 $\pm$ 1.04 & 2 &	0.330 & 266 & 0.91 & 0.116 $\pm$ 0.022 & 3.083 \\
7 & 65 & 197.97640 & -1.15380 & -22.99 & 0.85 $\pm$ 0.38 & 2 &	0.108 & 311 & 0.02 & 0.352 $\pm$ 0.100 & 2.339 \\
8 & 31 & 197.95769 & -1.17172 & -24.86 & 2.74 $\pm$ 1.20 & 3 &	0.090 & 196 & 0.40 & 0.080 $\pm$ 0.026 & 3.468 \\
9 & 38 & 197.95643 & -1.17477 & -24.28 & 7.51 $\pm$ 3.28 & 3 &	0.077 & -41 & 0.00 & 0.428 $\pm$ 0.030 & 3.748 \\
10 & 43 & 197.95238 & -1.18489 & -23.87 & 1.69 $\pm$ 0.74 & 2 &	0.049 & -157 & 0.37 & 0.158 $\pm$ 0.046 & 2.658 \\
11 & 37 & 197.94754 & -1.18383 & -22.75 & 0.74 $\pm$ 0.33 & 2 &	0.217 & -41 & 0.00 & 0.510 $\pm$ 0.096 & 2.489 \\
12 & 32 & 197.95437 & -1.17140 & -26.18 & 13.92 $\pm$ 6.08 & 3 &	0.149 & -62 & 0.95 & 0.037 $\pm$ 0.024 & 3.607 \\
13 & 14 & 197.94564 & -1.16432 & -23.64 & 1.58 $\pm$ 0.69 & 2 &	0.107 & 44 & 0.06 & 0.214 $\pm$ 0.049 & 2.696 \\
14 & 6 & 197.94071 & -1.16265 & -23.48 & 0.62 $\pm$ 0.28 & 2 &	0.361 & 88 & 0.03 & 0.444 $\pm$ 0.093 & 2.582 \\
15 & 2 & 197.93615 & -1.15561 & -22.81 & 0.93 $\pm$ 0.42 & 2 &	0.029 & 83 & 0.00 & 0.356 $\pm$ 0.095 & 2.130 \\
16 & 7 & 197.94359 & -1.15721 & -23.63 & 1.81 $\pm$ 0.79 & 2 &	0.138 & 0 & 0.00 & 0.257 $\pm$ 0.043 & 2.779 \\
17 & 20 & 197.95362 & -1.15161 & -23.04 & 0.46 $\pm$ 0.21 & 2 &	0.558 & 138 & 0.00 & 0.505 $\pm$ 0.092 & 2.285 \\
18 & 41 & 197.96484 & -1.15373 & -23.55 & 0.95 $\pm$ 0.42 & 2 &	0.174 & 198 & 0.00 & 0.410 $\pm$ 0.094 & 2.830 \\
19 & 35 & 197.96782 & -1.15578 & -25.13 & 2.27 $\pm$ 0.99 & 2 &	0.246 & 180 & 0.76 & 0.136 $\pm$ 0.026 & 3.321 \\
20 & - & 197.98916 & -1.15270 & -24.38 & 1.55 $\pm$ 0.19 & 2 &	0.040 & 98 & 0.00 & 0.171 $\pm$ 0.064 & 2.061 \\
21 & 69 & 197.97680 & -1.16486 & -22.56 & 0.52 $\pm$ 0.23 & 2 &	0.278 & 278 & 0.00 & 0.439 $\pm$ 0.107 & 2.504 \\
22 & 75 & 197.97635 & -1.18002 & -23.47 & 0.86 $\pm$ 0.38 & 2 &	0.301 & 193 & 0.00 & 0.626 $\pm$ 0.089 & 2.622 \\
23 & 61 & 197.96103 & -1.17819 & -23.19 & 0.57 $\pm$ 0.25 & 2 &	0.055 & 292 & 0.00 & 0.416 $\pm$ 0.092 & 2.822 \\
24 & 70 & 197.96526 & -1.19081 & -23.66 & 1.19 $\pm$ 0.52 & 2 &	0.245 & -124 & 0.09 & 0.290 $\pm$ 0.055 & 2.411 \\
25 & 60 & 197.96108 & -1.18797 & -23.64 & 0.85 $\pm$ 0.37 & 2 &	0.261 & -41 & 0.00 & 0.509 $\pm$ 0.092 & 2.469 \\
26 & 29 & 197.95675 & -1.17549 & -24.91 & 4.64 $\pm$ 2.02 & 3 &	0.116 & -26 & 0.45 & 0.120 $\pm$ 0.027 & 3.446 \\
27 & 42 & 197.95666 & -1.17153 & -24.94 & 3.36 $\pm$ 1.47 & 3 &	0.154 & -53 & 0.96 & 0.107 $\pm$ 0.029 & 3.660 \\
28 & 8 & 197.93336 & -1.18330 & -23.54 & 0.83 $\pm$ 0.37 & 2 &	0.617 & -35 & 0.00 & 0.588 $\pm$ 0.094 & 2.968 \\
29 & 28 & 197.95005 & -1.17060 & -23.63 & 0.69 $\pm$ 0.31 & 2 &	0.081 & 96 & 0.00 & 0.445 $\pm$ 0.099 & 3.060 \\
30 & 4 & 197.93953 & -1.16139 & -23.35 & 0.93 $\pm$ 0.41 & 2 &	0.265 & 70 & 0.03 & 0.370 $\pm$ 0.092 & 2.594 \\
\hline
\tablecomments{
Column (1): galaxy ID number used throughout this work.
Column (2): galaxy ID from \citet{halkola2006}.
Column (3): right ascension in degrees and decimal (J2000.0).
Column (4): declination in degrees and decimal (J2000.0).
Column (5): $K$-band galaxy magnitude derived from the apparent $r'$-band
magnitude and corrected as detailed in
\refsec{sec:discussion}.
Column (6): \re\ obtained with a curve of growth method and masking nearby
objects, see \refsec{sec:data.data.analysis.photometry}.
Column (7): quality of the \re\ determination. A value of 1 is only given to
the best fits. Values of 3 are assigned to objects with severe contamination.
Column (8): ellipticity determined with the method of moments, inside the
isophote of area $\pi \; R_{\rm e}^2$.
Column (9): position angle determined with the method of moments, inside the
isophote of area $\pi \; R_{\rm e}^2$.
Column (10): probability that the galaxy is a SR, see \reffig{fig:lambda.vs.epsilon}.
Column (11): $\lambda_R$ measured within the whole IFU field of view.
Column (12): Mean surface density of galaxies inside the circle centred on the
galaxy and containing its 3 closest neighbours.
}
\end{tabular}
\end{table*}

\section[]{Data}\label{sec:data}

\subsection{Data reduction}\label{sec:data.data.reduction}

We extracted the spectra using the standard ESO pipeline\footnote{http://www.eso.org/sci/software/pipelines/giraffe/giraf-pipe-recipes.html},
following the guidelines ESO offers\footnote{ftp://ftp.eso.org/pub/dfs/pipelines/giraffe/giraf-manual-2.8.7.pdf}.

Each morning the telescope produces a number of calibration frames, including
bias, lamp flats and arc lamp frames. To extract the spectra from the raw
images we used the closest calibration available.  The pipeline is organized
into 9 ``recipes'', distinct applications with a number of user configurable
parameters: we used the default values unless otherwise stated.

For each night we created a master bias out of the 5 raw frames provided. We
used the method \textit{ksigma} and the recipe \textit{gimasterbias}, with the
keywords \textit{ksigma.low} and \textit{ksigma.high} set to 3.0 to remove cosmic rays.

We then proceeded to ``fibre localization'', (tracing the spectra
on the chip). This is done using a set of 3 very high
SNR lamp flat frames, in the recipe \textit{gimasterflat}. At each spectral
pixel on the frame the recipe determines the locations, in the cross-dispersion
direction, of the light peaks corresponding to the centres of each fibre
signal. A curve is fitted to each profile, and is stored as the trace
shape.
We used the standard unweighted summation to extract the spectra (we set the keyword \textit{extraction.method} to \textit{SUM}).
We set to \textit{PROFILE+CURVE} the keyword
\textit{biasremoval.method}, as advised by ESO on the website, while the
keywords \textit{fibres.spectra} and \textit{fibres.nspectra} were modified to
take into account the occurence of both broken and unused fibres. The manual
indicates that, using \textit{SUM}, the contamination between neighboring spectra is less than
10\% of the counts. The recipe
also determines the pixel-to-pixel variation corrections and the fibre-to-fibre
transmission variations.
The wavelength calibration was done separately for each night using the recipe
\textit{giwavecalibration}. The resulting wavelength solution has 
an accuracy of $0.009 \pm 0.033 \mbox{\AA}$ and a resolution FWHM of $0.61
\pm 0.07 \mbox{\AA}$.
The science extraction was performed using \textit{giscience}. We set
the parameters \textit{biasremoval.method} to \textit{PROFILE+CURVE} and 
\textit{flat.apply} to \textit{TRUE}.

\subsection[]{Data analysis}

\subsubsection{Photometry}\label{sec:data.data.analysis.photometry}

We used \textit{g$'$} and \textit{r$'$} band Gemini imaging to create a
catalogue of all galaxies in the observed region of the
sky \citep{houghton2012}. We applied cuts at \textit{r$'$} = 22 and in the
related error ($\sigma_{r'}, \sigma_{g'} \; < \; 0.1$ mag). The resulting catalogue has been used to compute the
number density of galaxies (\refsec{sec:environment}), the cluster Luminosity
Function (LF)
(\refsec{subsec:sample.selection.effects}) and the cluster CMD
(\refsec{sec:red.sequence.selection}).
HST imaging was used to determine de Vaucouleurs \citep{devaucouleurs1953}
effective radii \re\ \citep[using the curve of
growth method of ][]{houghton2012} and ellipticities $\epsilon$, whenever
this was possible. In practice one galaxy (number 20 in \reftab{tab:sample})
lies partially outside the ACS field of view, and takes its photometric
parameters from the r$'$ Gemini image.

Since a large fraction of our sample is found in very dense regions, the
surface photometry is often contaminated by that of a neighbour. Consequently the
\re\ values in \reftab{tab:sample} include a quality flag Q, as in
\citet{houghton2012}.

Following \cite{cappellari2007} we adopted the method of moments to determine
ellipticities: after identifying the image isophotes we compute, for each of
them, the position angle of the major axis PA, the ellipticity $\epsilon$ and
the surface area $A$. The ellipticity of the k-th isophote $\epsilon_k$ is
defined by
\begin{equation}
(1 - \epsilon_k^2) \equiv \frac{\displaystyle\sum_{i \in \mathcal{I}_k} F_i y_i}
{\displaystyle\sum_{i \in \mathcal{I}_k} F_i x_i}
\end{equation}
where $F_i$ is the flux associated with the i-th pixel, and the 
coordinates $(x, y)$ are drawn from the galaxy centre, with the x-axis along the
photometric major axis. The sum is conducted on the set of
all pixels comprised in the k-th isophote. We associate to each isophote an
ellipse of area $A_k$ equal to the isophote area, ellipticity $\epsilon_k$ and
position angle $PA_k$, and associate to it a radius defined by $R_k \equiv
\sqrt{A_k / \pi}$.
\sauron\ and \atl\ based their classification on values computed at 1 \re.
We therefore define \epse\ as the value of $\epsilon_k$ computed within the
isophote of associated radius \re. The results are listed in \reftab{tab:sample}.
We find them to be robust against changes in \re, except for galaxy
number 9 (\reftab{tab:sample}), which exhibits peculiar photometry, having an
abrubt change in both $\epsilon$ and position angle at a radius of $\approx
0.5''$.

\subsubsection{Stellar kinematics}\label{sec:stellar.kinematics}

Stellar kinematics were extracted using \ppxf, a penalized
maximum likelihood algorithm developed by \cite{cappellari2004}. It fits the
line-of-sight velocity distribution (LOSVD) in pixel space, by convolving a
linear combination of stellar template spectra with an LOSVD expressed by the
truncated Gauss-Hermite series \citep{vandermarel1993, gerhard1993}:
\begin{equation}
\mathcal{L}(v)=\frac{e^{-(1/2)y^2}}{\sigma\sqrt{2\pi}}
    \left[ 1 + \sum_{m=3}^M h_m H_m(y) \right]
    \label{eq:losvd.gaussherm.dec}
\end{equation}
where $y = (v - V) / \sigma$ and the $H_m$ are Hermite polynomials. In practice
however, our SNR was mostly lower than that $(\approx 60)$ required to reliably
measure the weights $h_3$ and $h_4$ so we fitted a
Gaussian function, obtaining just $V$ and $\sigma$ in the above expression.
While \atl\ team used MILES stellar template library
\citep{sanchezblazquez2006}, its FWHM resolution of 2.54 \Angst\
\citep{beifiori2011} was lower than that of our data (see
\refsec{sec:data.data.reduction}), so we used the high resolution
version (R = 40000) of the ELODIE template library \citep{prugniel2001}, with FWHM
of 0.13 \Angst. The two libraries span similar regions in the age-metallicity
space; MILES reaches lower ages ($\approx 7$ Gyr vs $\approx 8$ Gyr) and
includes some old, metal poor stars ($\rm{Z} \approx 1/100 \rm{Z}_\odot$),
but these are not relevant when fitting ETGs, and the change
of library is unlikely to introduce any significant bias when compared to \atl\
measurements.
All ELODIE templates have a gap at $\lambda \approx 5414 \, \rm\AA$, so we cut the
galaxy spectra at 5300 \Angst.
For each galaxy we computed a weighted average (with sigma clipping rejection)
of all the 20 spectra, and fed it to \ppxf\ along with all the templates
available in ELODIE. This resulted in $\approx 15$ templates being selected for
each galaxy, and we use this subset to fit the individual fibre spectra of
the galaxy.
We used formal errors derived by \ppxf\ (we
did not exploit the penalizing functionality of the algorithm). These are
typically of the order of 15 \kms\ for V and 17 \kms\ for $\sigma$, but they
do not take into account the correlation introduced when log-rebinning.

Due to the high spectral resolution and low SNR, we decided not to subtract the sky, but rather to
fit it simultaneously with the stellar templates. Like \citet{weijmans2009}, we
provided \ppxf\ with all the simultaneous sky spectra and let the maximum
likelihood algorithm rescale them to best fit the data. 

\section[]{results}\label{sec:results}

\subsection{Kinematic maps}\label{sec:kinematic.maps}

\begin{figure*}
  \includegraphics[width=\textwidth]{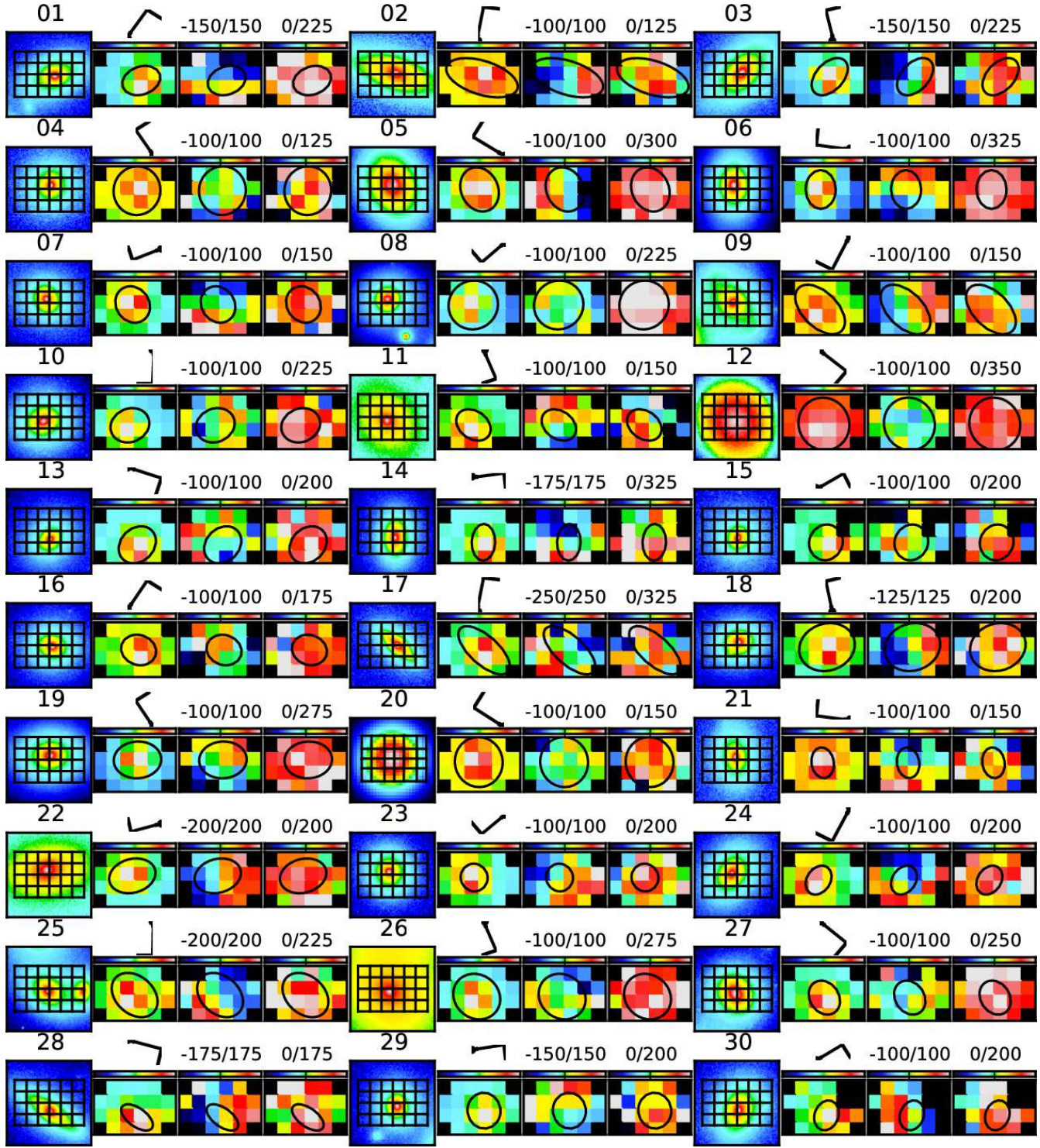}
  \caption{Kinematic maps of the Abell 1689 sample. Each horizontal set of four images
    depicts one of the 30 galaxies in the sample. The first plot shows HST
    photometry (apart from target 20). Superimposed is the FLAMES/GIRAFFE
    footprint.
    The second plot is the reconstructed image from VLT integral spectroscopy,
    where each square is a spaxel, corresponding to a lenslet in the
    instrument. Superimposed is an isophote at either \re, or the closest
    fraction that fits into the IFU footprint. The four black corners
    correspond to
    unused ``dead'' corners, while other black spaxels (seen in 11, 15 and 30)
    correspond to broken or unused fibres.
    The third and fourth plots depict the kinematic maps: velocity and velocity
    dispersion. The black compass arrows show North and East. The
    colorbar limits are given in \kms.}
    \label{fig:kinematic.maps}
\end{figure*}

The results of the kinematic extraction can be seen on
\reffig{fig:kinematic.maps}. There are four frames for each galaxy, from left
to right: high
resolution photometry (from either HST or GEMINI); low resolution GIRAFFE
spectrograph photometry; the extracted velocity map and the extracted velocity dispersion
map. Above each galaxy we give the ID number; the
celestial orientation is given by the black compass arrows (N and E). Corner
spaxels and spaxels corresponding to broken/unused optic fibres are depicted in black.
Although the spatial resolution is low, rotation can be clearly seen in
some galaxies, while no such features are seen on others. 

We cannot detect kinematically decoupled cores (KDCs) and double
$\sigma$ peaks (2-$\sigma$) as in \citet{krajnovic2011}, because our spatial
resolution is too coarse.
If we try to detect SRs from the velocity maps by eye, we identify at
most six: these are galaxies 4, 8, 12, 20, 26 and
27. The overall fraction of SRs in the sample would then be 0.20, in
line with what was found in the Virgo core \citep{cappellari2011b}.
However we are subject to contamination from face on discs appearing as SRs,
which increases \fsr.

We also highlight five more
objects which, despite exhibiting large scale rotation, have misaligned
kinematic axes, a
feature more common in SRs than in FRs \citep{krajnovic2011}: these are
galaxies 1,
3, 5, 9, 17 and 25. Galaxies 3 and 17 have very high ellipticities, and are thus unlikely
to be SRs. Galaxy 5 has high velocity dispersion, and also contains an inner
disc (R = 1.5 kpc) in the HST imaging.

\subsection{$\lambda_R$ and kinematic classification.}\label{sec:slow.fast}

\citet{emsellem2007} introduced the estimator $\lambda_R$ to measure the projected specific angular
momentum of galaxies and \citet{emsellem2011} further show how the combination
of $\lambda_R$ and ellipticity $\epsilon$ conveniently
captures the kinematic boundary between SRs and FRs.
$\lambda_R$ is defined as
\begin{equation}\label{eq:lambda.def}
  \lambda_R (I) \equiv
  \frac{\displaystyle\sum_{i \in \mathcal{I}}
                     F_i R_i |V_i|}{\displaystyle\sum_{i \in \mathcal{I}}
                     F_i R_i \sqrt{V_i^2 + \sigma_i^2}}
\end{equation}
where $F_i$, $R_i$, $V_i$ and $\sigma_i$ are the flux, distance from the galaxy
centre, velocity and velocity dispersion of the i-th spaxel; the sum is
conducted over all spaxels inside some subset $I$ of the IFU
footprint. \citet{emsellem2007, emsellem2011} define \lambdae\ as
the value of $\lambda_R$ computed inside the ellipse of area $\pi R_{\rm e}^2$
(see \refsec{sec:data.data.analysis.photometry}). In our
study however that ellipse may either not comprise enough
spaxels to reliably measure \lambdae, or be too large to fit inside the IFU
footprint.
Therefore we used existing \sauron\ data to estimate how our particular
observing setup affects the measured value of $\lambda_R$.

\subsubsection{Effect of pixelisation on $\lambda_R$.}\label{sec:effect.pixel}

The original \sauron\ sample covers a wide range of ETGs types \citep{dezeeuw2002}, and
its data is publicly available\footnote{http://www.strw.leidenuniv.nl/sauron/}.
We use it to simulate observations with FLAMES/GIRAFFE, in order to
determine how distance and reduced spatial resolution affect measurements of
$\lambda_R$. For each galaxy we created a kinematic model using \texttt
kinemetry\footnote{The IDL KINEMETRY routine can be found at
http://www.eso.org/~dkrajnov/idl/} \citep{krajnovic2006}; each model was
then projected to the distance of Abell 1689 and convolved with a seeing of
0.8$''$, before being ``observed'' with FLAMES/GIRAFFE.
We created 10000
realizations of each model, adding Gaussian errors
of 15 \kms\ and 17 \kms\ for $V$ and $\sigma$ respectively
(\refsec{sec:stellar.kinematics}), and proceeded to measure $\lambda_R$ for
each of them.

In their $\lambda_R$ vs $\epsilon$ diagram, \citet{emsellem2007,
emsellem2011} plot values computed on the same \sauron\ spectrograph images, at
the same spatial scale of 1 \re.
For the small galaxies in our sample however, \re\ covers just a few pixels
whereas the large galaxies have \re\ larger than the field of view of the IFU.
For this reason we cannot follow the \atl\ prescription precisely. 
We therefore introduced $\lambda_R(IFU)$, defined as the value of $\lambda_R$
computed using all the 
available spaxels in the IFU field of view and show through simulation 
of the SAURON results that it is a satisfactory proxy for \lambdae.

\begin{figure}
  \includegraphics[width=\columnwidth]{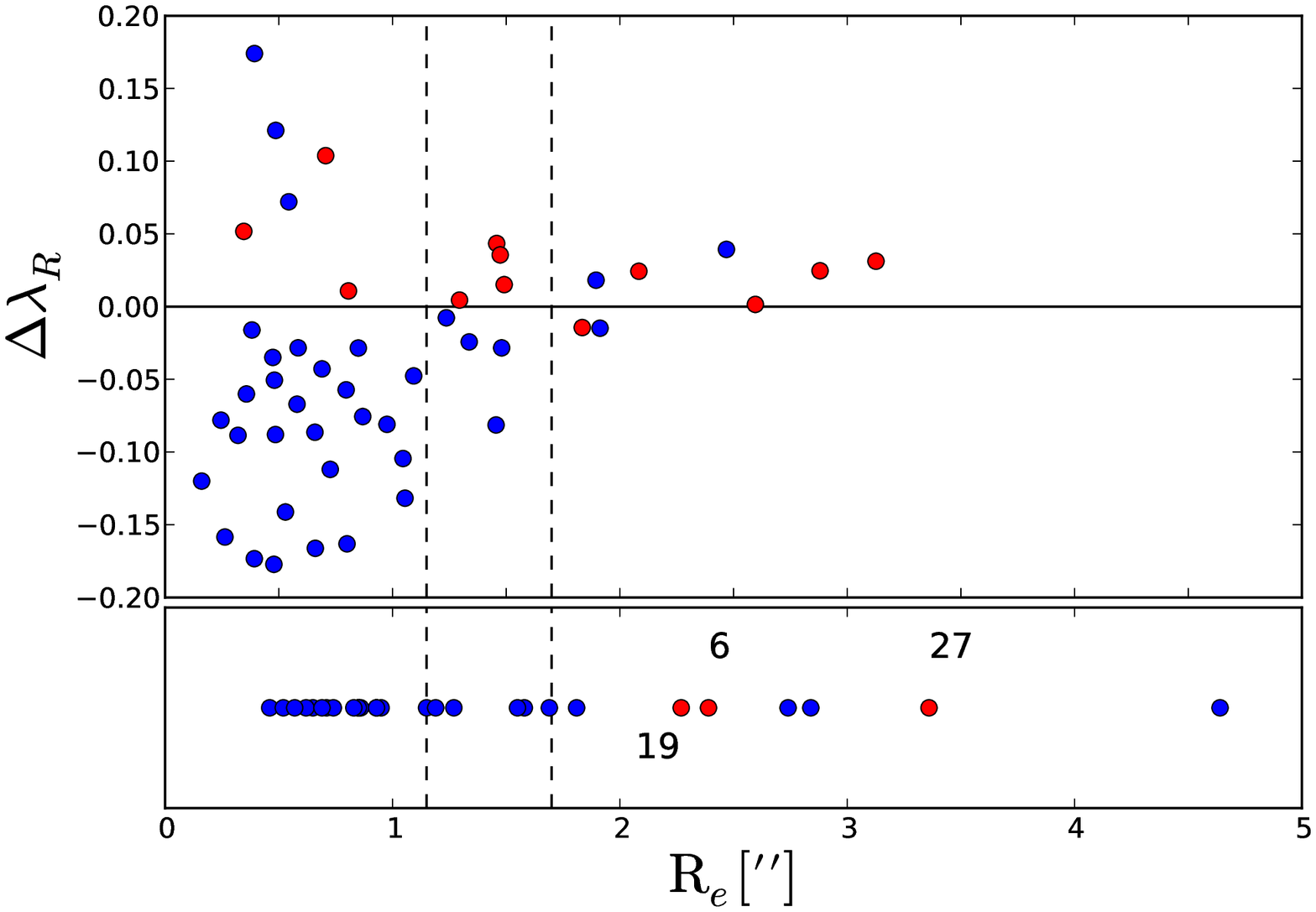}
  \caption{Simulated observation of \sauron\ data with FLAMES/GIRAFFE, at
  $z=0.183$. $\Delta \lambda_R$ is the difference between $\lambda_R(IFU)$
  (computed from simulated observations of \sauron\ data with GIRAFFE) and the
  value \lambdae\ given in \citet{emsellem2007}, plotted against \re.
  Slow/fast rotators are denoted by red/blue
  dots, and classified as in \citet{emsellem2007} (upper panel). In the lower
  panel we show the values of \re\ for our sample of galaxies in Abell 1689,
  where the symbol colour indicates SRs (red) and FRs (blue), according to a
  classification done using the $\lambda_R$-$\epsilon$ plot
  (\reffig{fig:lambda.vs.epsilon}), as described in
  \refsec{sec:lambda.measurements}. Dashed vertical lines define regions
  over which we estimate biases and systematic errors
  (see \refsec{sec:effect.pixel}).}\label{fig:delta.lambda.vs.re}
\end{figure}

\reffig{fig:delta.lambda.vs.re} shows $\Delta \lambda_R$ plotted against \re,
 where $\Delta \lambda_R$ is defined as the
difference between $\lambda_R(IFU)$ and the value of \lambdae\ of
\citet{emsellem2007}. We can use this information to
determine the correction and the uncertainty that we need to apply to
$\lambda_R(IFU)$ to obtain \lambdae. It is clear how our ability to
recover the true value of \lambdae\ improves with increasing \re. To
make use of this information we separate the sample into three groups, based on \re\ (the divisions are at
\re\ values of 1.15$''$ and 1.70$''$ which naturally divide the Abell 1689
sample and are shown as vertical dashed lines in \reffig{fig:delta.lambda.vs.re}).
We find the following biases (mean offset) and systematic errors (dispersion): for galaxies with
\re$< 1.15''$ $\Delta \lambda_R = -0.06 \pm 0.09$; for galaxies with
$1.15''\leq$\re$<1.7''$, $\Delta \lambda_R = -0.01 \pm 0.04$, and for the
remainder, $\Delta \lambda_R = 0.01 \pm 0.02$.

\begin{figure}
  \includegraphics[width=\columnwidth]{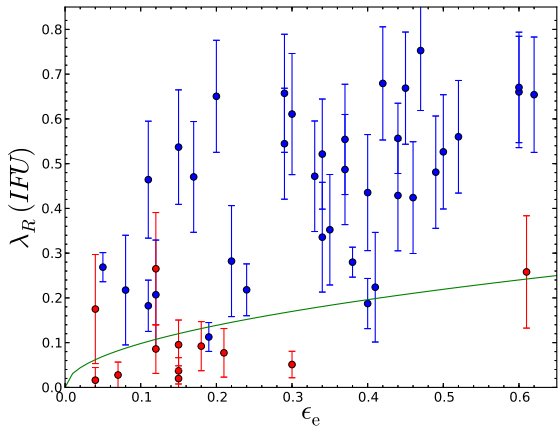}
  \caption{$\lambda_R(IFU)$ vs \epse\ for our simulated FLAMES/GIRAFFE
  observations of the \sauron\ sample of galaxies. The green lines, which
  separates SRs (below it) from FRs, has equation $\lambda_R = 0.31 \;
  \sqrt\epsilon$. Red and blue dots denote SRs and FRs respectively, according
  to the original \sauron\ classification \citep{emsellem2007}. While $\lambda_R(IFU)$ has
  been measured on the redshifted and resampled data, the values of $\epsilon$
  on the x axis
  are the original values published in \citet{emsellem2007}. Despite the latter
  being measured on much higher resolution than $\lambda_R(IFU)$, and at a
  different radius, the impact on the classification is low. Misclassified
  galaxies correspond either to red dots above the green line and blue dots
  below it.}\label{fig:lambda.vs.epsilon.sauron}
\end{figure}

We corrected $\lambda_R(IFU)$ according to the biases measured, summing the
systematic errors in quadrature to the random errors. This correction takes
into account both the different spatial scale between $\lambda_R(IFU)$ and
\lambdae\ and the different spatial resolution between $\lambda_R(IFU)$ and
\epse.
In \reffig{fig:lambda.vs.epsilon.sauron} we plot simulated values of
$\lambda_R(IFU)$ against published values of \epse\ \citep[from ][]{emsellem2007}.
Despite the aforementioned differences, there is little (10\%)
misclassification in our diagram, especially at high values of \re.
We can calculate the probability distribution for the number of SRs 
\citep[galaxies below the line defined by $0.31 \times \sqrt\epsilon$
and the green line in \reffig{fig:lambda.vs.epsilon.sauron},][]{emsellem2011}.
This is most easily done with a Monte Carlo approach.
For each galaxy we assume Gaussian errors in $\lambda_R$, truncated so that $0
\leq \lambda_R \leq 1$ and sample 100000 times.
The resulting probability distribution is Gaussian-like and we find 12.3 $\pm$
1.7 slow rotators, where the true value is 12. This
justifies both our choice of $\lambda_R(IFU)$ to substitute for \lambdae, and
the use of $\epsilon$ computed at a different resolution and radius than
$\lambda_R(IFU)$.

\subsubsection{$\lambda_R$ measurements and the statistical calculation of \fsr}\label{sec:lambda.measurements}

\begin{figure}
  \includegraphics[width=\columnwidth]{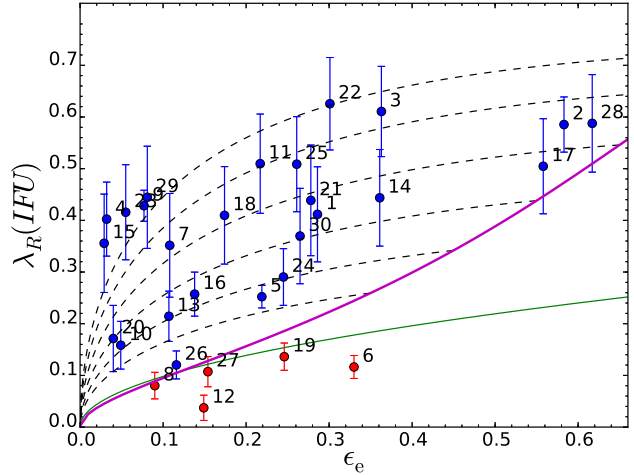}
  \caption{$\lambda_R(IFU)$ vs \epse\ for all target galaxies in Abell 1689. The green line
  has equation $\lambda_R = 0.31 \; \sqrt\epsilon$ and separates fast rotators
  (blue dots above it) from slow rotators (red dots below it).
  Error bars are dominated by the systematic error
  (\refsec{sec:lambda.measurements}). Notice that we corrected the measured
  value of $\lambda_R(IFU)$ by subtracting -0.06, -0.01 and 0.01,
  depending on \re\ for each galaxy (see again \refsec{sec:lambda.measurements}).
  The solid magenta line represents the edge-on view of
  axisymmetric galaxies with $\beta = 0.70 \times \epsilon$, while the black
  dashed lines represent the trajectories of 6 of these galaxies (with
  $\epsilon = 0.85, 0.75, ..., 0.35$) as their viewing angle goes from edge-on (on
  the magenta line) to face on (towards the origin). For more information on
  how these models were constructed see \citet{emsellem2011}}\label{fig:lambda.vs.epsilon}
\end{figure}

In \reffig{fig:lambda.vs.epsilon} we show the $\lambda_R(IFU)$ vs \epse\
plot for our Abell 1689 data.  Given the simulation in the previous section,
the values of $\lambda_R(IFU)$
have been corrected by -0.06, -0.01 and 0.01 for galaxies with \re$<1.15''$,
$1.15''\leq$\re$<1.7''$ and \re$\geq 1.7''$ respectively. The errors
include both the formal random error (from \ppxf) and the systematic error
(0.09, 0.04 and 0.02 for the three ranges of \re\ from the previous simulations).
Given these errors we can calculate the probability distribution for the number
of SRs, as done previously for the simulated \sauron\ data. We find 4.5 $\pm$
1.0 slow rotators, corresponding to \fsr=0.15 $\pm$ 0.03.
Galaxy number 9, which has peculiar photometry and an uncertain value of \epse,
has no effect on the result, because its value of $\lambda_R(IFU)$ is greater
than $\approx$ 0.25 (the maximum allowed for any SR) by more than 3$\sigma$.

\citet{emsellem2007} warn about using only $\lambda_R$ to assign a galaxy to
either the slow or fast rotator class. The discrepancy between the ``by eye''
classification and the classification here bolsters that warning. However, when studying
galaxies beyond the local Universe, such a detailed analysis as was done
by the \atl\ team is unfeasable. We are thus forced to rely on a statistical
approach.

\subsection{Environment density}\label{sec:environment}

\begin{figure}
  \includegraphics[width=\columnwidth]{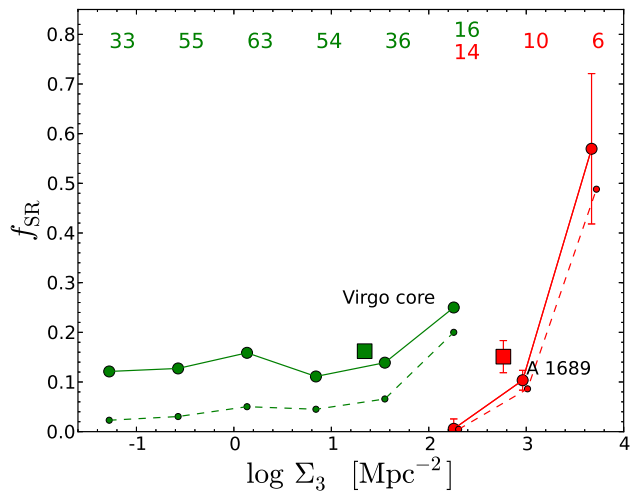}
  \caption{Fraction of slow rotators $f_{\rm SR}$ over the ETGs population in
  \atl, including Virgo cluster (green circles, solid green line), as given in \citet{cappellari2011b}, and fraction of slow
  rotators in our sample of Abell 1689 galaxies (red circles, solid red line). Numbers at the
  top are the total number of galaxies in that bin, with the same color code.
  The error bars for the Abell 1689 points represent the uncertainty in the
  slow rotators classification, as estimated in \refsec{sec:lambda.measurements}.
  The green square is the value of $f_{\rm SR}$ that we measure resampling
  Virgo using our sample luminosity function. The error bars are smaller than
  the marker size. The red square
  is the average fraction of slow rotators found in our
  sample. The lower, smaller circles and dashed lines are the fractions computed with respect to the
  total cluster population, for \atl\ (green) and Abell 1689 (red); this is an estimate based on
  spirals and blue ellipticals counts.}
  \label{fig:frac.vs.density}
\end{figure}

For each galaxy in the sample we computed the local environment density following
\citet{cappellari2011b}. We defined $\Sigma_3$ as the number density inside
the circular area centred around the target galaxy and
encompassing three other galaxies. Density estimates were done using only valid targets
in the catalogue of \refsec{subsec:sample.selection.effects}. We
applied a constant field correction of 0.49 gal arcmin$^{-2}$, measured
averaging data from one hundred 1 arcmin$^2$ fields from SDSS DR7 \citep{abazajian2009},
without correcting for cluster/groups contamination. For
comparison, the minimum value found in our sample is 3.83 gal arcmin$^{-2}$.
In \reffig{fig:frac.vs.density} we show \fsr\ versus $\Sigma_3$; we compare it
to the results of the \atl\ survey, and in particular to \fsr\ \citep{cappellari2011b}. The Virgo core
corresponds to the densest bin in \atl, with $f_{\rm SR} = 0.25$,
double that typically found in less dense environments ($f_{\rm SR} \approx
0.12$). We probe environments with values of $\rm{log_{10}} \Sigma_3$
between $2.06$ and $3.75$: the minimum is comparable to
the core of Virgo, and the maximum is 1.7 dex higher. In
this sense our work starts exactly where \atl\ finished. We find a
sharp increase in \fsr\ with projected density, ranging from \fsr$=0.01$ in the
least dense environment to \fsr$=0.58$ in the densest environment. Errors due to
misclassification, albeit large, show that the densest bin in Abell 1689 has a
higher fraction of SRs than Virgo core (\reffig{fig:frac.vs.density}). The
intermediate bin has a value of \fsr\ compatible, within the errors, with both
the field-group environments and overall Virgo cluster value but is
however lower than the Virgo core.
\fsr\ in the least dense bin is lower than \atl\ field and group
values.

However, considering the whole Abell 1689 sample, we find for an average value
of $\rm{log_{10}} \Sigma_3 = 2.77$ that the SR fraction is $0.15 \pm 0.03$ (red
square
in \reffig{fig:frac.vs.density}), which is the same as the overall SR fraction
in the Virgo cluster, when sampled in the same way (green square).
Furthermore, both values are similar to the field and group samples in \atl,
suggesting little to no difference in \fsr\ when it is averaged over the whole
cluster.

\section[]{Discussion}\label{sec:discussion}

\subsection{Sample selection effects}\label{subsec:sample.selection.effects}

In order to assess the robustness of our result, it is important to study the
relation between the sample of 30 galaxies and its parent population. Our
sample selection, limited by both observing and instrument constraints,
biases our study in different ways. In this section we discuss how these effects
change \fsr.

\begin{figure}
  \includegraphics[width=\columnwidth]{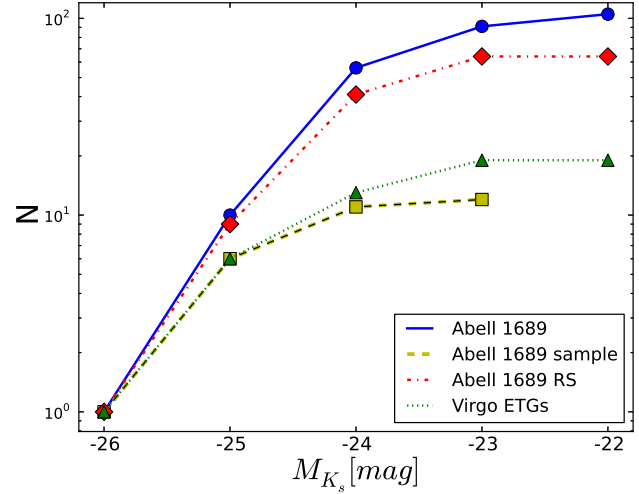}
  \caption{Abell 1689 (blue circles) and sample (yellow squares) K-band LFs,
  obtained from GEMINI/GMOS g$'$ band and r$'$ band imaging. r$'$ band
  magnitudes have been converted to $\rm M_{\rm K_{\rm s}}$ band as described in
  \refsec{sec:discussion}. Red diamonds trace the cluster RS, as
  determined by \citet{houghton2012}. Green triangles represent Virgo ETGs
  LF, from \atl\ survey (data available at
  www-astro.physics.ox.ac.uk/atlas3d/)}\label{fig:lum.func}
\end{figure}

Having in mind the \atl\ study of the Virgo cluster as a point of comparison, we determined the
Abell 1689 K-band LF. We took the r$'$-band catalogue
(\refsec{sec:data.data.analysis.photometry}) and, following
\citet{houghton2012}, applied a k-correction to the GMOS $r'$-band magnitudes.
The results have been converted to K$_s$ band (and Vega system), using
\citet{maraston2005} models, where we assumed an age of 10.4 Gyr
\citep{houghton2012} and passive evolution. We finally applied a  cut at
$\textrm{M}_\textrm{K} = -21.5\; \rm{mag}$, thus matching \atl\ parent sample selection.
The result is shown in \reffig{fig:lum.func} (blue circles), where we compare
it with the cluster RS \citep[as determined by][ red diamonds]{houghton2012}
and our sample (yellow squares). The Virgo ETG LF is also plotted
(green triangles).

\begin{figure}
  \includegraphics[width=\columnwidth]{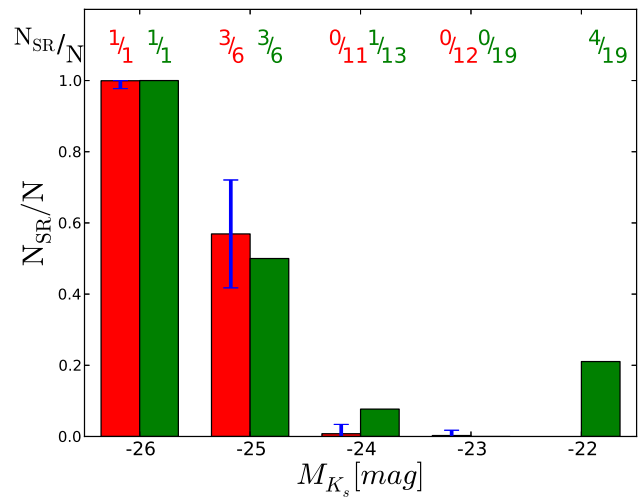}
  \caption{Fraction of SRs for Abell 1689 sample
  (red histogram) and in Virgo ETGs (green), as a function of magnitude. The number of
  SRs and the total number of galaxies in each magnitude bin are given at the
  top. For Abell 1689 the number of SRs has been rounded off.
  (\refsec{sec:lambda.measurements}).}
  \label{fig:lum.func.frac}
\end{figure}

Knowing the K-band magnitudes of our sample, we can show \fsr\ as a
function of magnitude. In \reffig{fig:lum.func.frac}, the value of \fsr\
observed in our sample (red) is compared to the
fraction of SRs over the ETGs population of Virgo (green). The two are,
within the errors, remarkably similar; however we do not reach magnitudes
beyond $\approx -23\; \rm{mag}$ to probe the faint SRs.

\subsubsection{Red sequence bias}\label{sec:red.sequence.selection}

Our sample falls entirely on the red sequence (RS), a property that was not sought after.
We know that the RS does not necessarily trace the
morphological ETGs population, as it can include red spirals and
omit blue ellipticals.
How many ETGs lying off the RS have we left out of our sample? A rich,
relaxed cluster like Abell 1689 comprises a very small fraction of spirals,
particularly in the core. In fact the ratio between the
RS LF and the cluster LF goes from 1 at the bright end to 0.70 at \MK = -22.5
$\rm{mag}$.
This means that including 10 ``blue'' galaxies in the faintest magnitude
bins would remove the RS bias, leaving us with a color ``fair'' sample.
These faint galaxies are more likely to be FRs (\reffig{fig:lum.func.frac}), so the bias introduced by
selecting galaxies on the RS leads us to overestimate \fsr.
In fact, if we assume that these ten galaxies are all FRs, and that they are distributed
spatially much like the observed targets, we can determine the kinematic
morphology-density relation for an unbiased sample (with respect to color) which we show in
\reffig{fig:frac.vs.density} (smaller red dots and red dashed line), with the
relation from \atl\ (smaller green dots and green dashed line).
The good overlap between ETGs and the RS in Abell 1689 causes \fsr\ to stay the
same, whether the fraction is computed against the RS or overall
galaxy population. This is not true in a less relaxed, spiral rich cluster like
Virgo (\reffig{fig:frac.vs.density}).

\subsubsection{Magnitude selection}\label{sec:magnitude.selection}

We know that in Virgo, \fsr\ varies as a function of \MK\ (\reffig{fig:lum.func.frac}),
and that the LFs of Virgo and Abell 1689 are different, in that Virgo is
relatively richer in brighter objects (\reffig{fig:lum.func}). Since our sample
is not fully representative of the Abell 1689 population, what bias does this
introduce in the measured value of \fsr?
A rigorous answer to this question is impossible, because we do not know if the
SR LF varies as a function of redshift and/or environment. In particular,
Virgo is a small and dynamically young cluster, whereas Abell 1689 is a massive,
relaxed system. However, using a simulation, we can estimate what SR
fraction we would measure in Virgo with the same selection effects present in
our Abell 1689 sample.

Let us assume that the SR fraction as a function of magnitude is the same in
Abell 1689 and Virgo (reasonable given \reffig{fig:lum.func.frac}) and that the galaxies were
selected only on their magnitude and no other properties (not true, as
discussed in \refsec{sec:sample.selection}, but reasonable given the substitutions required to
comply with proximity constraints).
Using the actual Abell 1689 LF, we drew random subsamples from the \atl\ Virgo
ETG population. These samples yielded \fsr\ = $0.16 \pm 0.01$, in agreement
with the actual Virgo value of $0.16$. Thus despite being biased towards
brighter galaxies, we should measure the same \fsr; this is because we sample down to, but not including, the faintest magnitude
bin of \atl, where \fsr\ suddenly increases.

\subsubsection{Other factors}

We remark that the distribution of projected ellipticities \epse\ of
our sample is different from that observed in Virgo. Since the former is richer
in round objects, and since SRs generally appear rounder, it follows
that our sample could be biased towards higher values of \fsr. It may well be
that the clusters $\epsilon$ distributions are different, in which case a
higher fraction of round objects may increase \fsr.

Another possible source of bias is the intrinsic shape of Abell 1689: according
to \citet{oguri2005} and \citet{corless2009}, Abell 1689 is elongated along the line
of sight, so that its measured $\Sigma_3$ is higher than what we would observe
from another point of view. If the cluster length along the line of sight
direction were $\gamma$ times longer than the diameter of the sky projection, then
the value of $\Sigma_3$ observed would be $\approx \gamma$ times the unbiased
value. Since the maximum reasonable value of $\gamma$ is $\approx 3$, in
\reffig{fig:frac.vs.density} $\log \Sigma_3$ is overestimated by at most $\approx 0.5$, which does not
significantly affect
our results.

Finally we remark that the corrections to $\lambda_R(IFU)$ that we derived in
\refsec{sec:effect.pixel} increase \fsr; had we applied no correction, we would
have 3.8 $\pm$ 1.0 SRs, so an even lower value of \fsr.

\subsection{General remarks}

Abell 1689 has a higher average density than Virgo, but the same value of \fsr.
Inside the cluster, in agreement with the findings of \citet{cappellari2011b}, \fsr\ rises with projected
density. In the least dense region \fsr\ is smaller than the \atl\ field/group value. Given the low number of
galaxies per bin, we cannot rigorously claim that this is representative.
However, a similar ``depletion'' is observed in the outskirts of
Virgo cluster \citep{cappellari2011b}.
One explanation could be that massive SRs are driven by
dynamical friction towards the centre of the cluster.
If these were originally distributed in the cluster like other galaxies,
dynamical friction would reduce their orbital velocity and radius.
Since this process is more effective on more massive galaxies, it would
concentrate SRs (more massive on average) with respect to other galaxies.

\section[]{Summary}\label{sec:summary}

We demonstrated the use of FLAMES/GIRAFFE in IFU mode to perform a
survey of 30 galaxies in Abell 1689 at $z=0.183$. The data has sufficient quality and spatial resolution to classify the majority of
targets as either SRs or FRs. In summary:

\begin{enumerate}
  \item  we find, in agreement with \atl\ results, that SRs populate the high luminosity end
of the LF; the SR
LFs measured from the Virgo \atl\ sample and our Abell 1689 sample are identical down to
$\rm{M_K} = -23$ mag.
  \item the fraction of
slow rotators in our sample is \fsr=$0.15 \pm 0.03$.
If we apply the same selection criteria to all Virgo galaxies in \atl, we find
the same fraction (assuming that the distribution of SRs with magnitude is the
same in both clusters). This indicates that \fsr\ is not affected by the
average number density of the cluster. Both Abell 1689 and Virgo average \fsr\
are in line with the \atl\ value for field and group environments.
  \item the fraction of SRs increases towards the denser, central region of the
cluster. This is in agreement with what was found in Virgo, where SRs
concentrate in the cluster core. This could be a consequence of dynamical
friction, as SRs dominate the high mass end of the galaxy population.
\end{enumerate}

It is important to expand this study, both to further study Abell 1689
down to lower luminosities and increase the number of observed clusters, to
quantify the scatter in \fsr.

\section*{Acknowledgments}

We thank Michele Cappellari and John Magorrian for their helpful comments and
suggestions. We are particularly grateful to the referee Eric Emsellem, whose
comments significantly improved this paper.
We wish to express our thanks to Jonathan Smoker, who was
instrument scientist for FLAMES when we took the observations.
FDE acknowledges support from the Physics Department,
University of Oxford and travel support from Merton College, Oxford.
RLD acknowledges travel and computer grants from Christ Church College, Oxford, and
support from the Royal Society in the form of a Wolfson Merit Award
502011.K502/jd. RLD also acknowledges the support of the ESO Visitor Programme
which funded a 3-month stay in 2010.

EDB was supported by the grants CPDA089220/08 and 60A02-5934/09 of Padua
University, and ASI-INAF I/009/10/0 of Italian Space Agency, and by a grant of
Accademia dei Lincei and Royal Society. EDB acknowledges the Sub-department
of Astrophysics, Department of Physics, University of Oxford and Christ Church
College for the hospitality while this paper was in progress.

This paper is based on observations made with ESO Telescopes at the Paranal
Observatory under programme ID 083.B-0681(A) and the NASA/ESA Hubble Space
Telescope (proposal ID 9289), obtained from the data archive at the Space
Telescope Science Institute. STScI is operated by the Association of
Universities for Research in Astronomy, Inc. under NASA contract NAS 5-26555.
This work also used data obtained at the Gemini Observatory
(proposal IDs GN-2001B-Q-10 \& GN-2003B-DD-3), which is operated by the
Association of Universities for Research in Astronomy, Inc., under a
cooperative agreement with the NSF on behalf of the Gemini partnership: the
National Science Foundation (United States), the Science and Technology
Facilities Council (United Kingdom), the National Research Council (Canada),
CONICYT (Chile), the Australian Research Council (Australia), Ministrio da
Cincia, Tecnologia e Inovao (Brazil) and Ministerio de Ciencia, Tecnologia e
Innovacion Productiva (Argentina).

\bibliographystyle{mn2e}

\label{lastpage}

\end{document}